# Suppression of X-Ray-Induced Radiation Damage to Biomolecules in Aqueous Environments by Immediate Intermolecular Decay of Inner-Shell Vacancies

*Andreas Hans[†*], Philipp Schmidt[†,‡], Catmarna Küstner-Wetekam[†], Florian Trinter[$,¥],*

*Sascha Deinert[$], Dana Bloß[†], Johannes H. Viehmann[†], Rebecca Schaf[†], Miriam Gerstel[†],*

*Clara M. Saak[§], Jens Buck[$], Stephan Klumpp[$], Gregor Hartmann[†,¶],*

*Lorenz S. Cederbaum[♯], Nikolai V. Kryzhevoi[♯*] and André Knie[†]*

[†]Institut für Physik und CINSaT, Universität Kassel, Heinrich-Plett-Straße 40, 34132 Kassel, Germany

[‡]European XFEL, Holzkoppel 4, 22869 Schenefeld, Germany

[$]Deutsches Elektronen-Synchrotron (DESY), Notkestraße 85, 22607 Hamburg, Germany

[¥]Molecular Physics, Fritz-Haber-Institut der Max-Planck-Gesellschaft, Faradayweg 4, 14195 Berlin, Germany

[§]Molecular and Condensed Matter Physics Division, Department of Physics and Astronomy, Uppsala University, Box 516, 75120 Uppsala, Sweden

[¶]Helmholtz-Zentrum Berlin (HZB), Albert-Einstein-Straße 15, 12489 Berlin, Germany

[♯]Theoretische Chemie, Physikalisch-Chemisches Institut, Universität Heidelberg, Im Neuenheimer Feld 229, 69120 Heidelberg, Germany

AUTHOR INFORMATION

Corresponding Authors

*E-mail: hans@physik.uni-kassel.de

*E-mail: nikolai.kryzhevoi@pci.uni-heidelberg.de

ABSTRACT

The predominant reason for the damaging power of high-energy radiation is multiple ionization of a molecule, either direct or via the decay of highly excited intermediates, as e.g., in the case of X-ray irradiation. Consequently, the molecule is irreparably damaged by the subsequent fragmentation in a Coulomb explosion. In an aqueous environment, however, it has been observed that irradiated molecules may be saved from fragmentation presumably by charge and energy dissipation mechanisms. Here, we show that the protective effect of the environment sets in even earlier than hitherto expected, namely immediately after single inner-shell ionization. By combining coincidence measurements of the fragmentation of X-ray-irradiated microsolvated pyrimidine molecules with theoretical calculations, we identify direct intermolecular electronic decay as the protective mechanism, outrunning the usually dominant Auger decay. Our results demonstrate that such processes play a key role in charge delocalization and have to be considered in investigations and models on high-energy radiation damage in realistic environments.

# MAIN TEXT

The exposure to ionizing radiation is known to have severe consequences to living organisms. Depending on the radiation dose, the organism may suffer from cytotoxic effects ranging from enhanced cancer risk to radiation sickness. Nonetheless the macroscopic symptoms, the radiation damage itself happens on a molecular level[1].

As a common assumption, about two-thirds of such damage is an *indirect* and purely environmental effect caused by secondary low-energy electrons and radicals[1–4] originating in radiolysis of water monomers surrounding biomolecules. The rest results from *direct* deposition of energy in biomolecules. It is tempting to assume that the indirect damage owes its predominance to simply the large amount of water in biological tissue. The situation is, however, more complicated, and recent works suggest that the contribution from direct and indirect processes is still not well understood[3; 5].

As evidenced by recent ion-impact experiments on several hydrated biomolecules[6–10], an aqueous environment is able to significantly suppress the direct damage and thus play an "unusual" role of a damage protector. This striking protective effect was ascribed to the environmental ability to absorb and dissipate the energy transferred to biomolecules in collisions[7; 11]. If the radiation damage was induced by multiple ionization, charge redistribution from localized states to delocalized ones was also considered among the potential protection mechanisms. In the present study, we demonstrate that the water environment starts to protect biomolecules exposed to radiation even before ultrafast autoionization converts these molecules into highly unstable multiply charged systems.

In case of X-ray element-specific irradiation, the radiobiologically most relevant process is inner-shell ionization of one of the constituting atoms of a biomolecule, forming highly excited states with vacancies in a core orbital. Such highly excited species relax predominantly via Auger decay, in which one electron of the molecular valence orbitals fills the core vacancy and another valence electron is released, resulting in dissociative multiply charged states. The consequences, namely inevitable destruction by fragmentation, are well investigated for isolated molecules[12–20].

We investigated the X-ray-induced fragmentation of microsolvated pyrimidine in a photoelectron-photoion-photoion coincidence (PEPIPICO) experiment (see Methods section). Being one of the building blocks of larger biomolecules such as nucleic and amino acids, pyrimidine ($C_4H_4N_2$) serves as a prototype system for many fundamental investigations[15; 20]. Its photochemistry is especially relevant for various medical applications, in which halogenated pyrimidines are used as radiosensitizers[21]. The core ionization of isolated pyrimidine and its derivatives is known to cause the fragmentation of the molecule into two or more ionic fragments[14; 15]. Thus, no parent ions or doubly charged molecular fragments are observed in coincidence with core electrons, which is in agreement with the present work.

In contrast, for microsolvated pyrimidine we observe intact parent ions in coincidence with water cluster ions and carbon photoelectrons. The corresponding ion-ion coincidence map is shown in Fig. 1a. The mass-to-charge ratio of the pyrimidine parent ion is m/z = 80. Prominent features appear in the coincidence map on the m/z = 80 diagonal or to the upper left of it, i.e., for pairs of singly charged ions whose sum of masses is 80 atomic units or less. These features mainly originate from ionization of isolated pyrimidine molecules, which occur abundantly in the target (see Experimental Methods section).

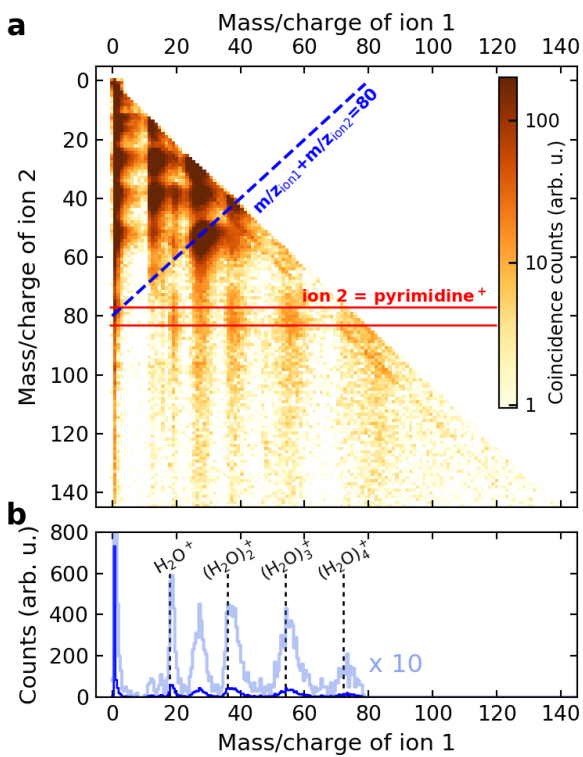

**Figure 1.** Photoelectron-photoion-photoion coincidence spectra from a target jet containing both isolated and microsolvated pyrimidine molecules, ionized with soft X-rays of 300 eV photon energy. **a**, Ion-ion map for coincidences of a carbon core photoelectron and two ions. **b**, Spectrum of ion 1, selected for ion 2 being the pyrimidine parent ion (dark blue trace) and tenfold magnified (light blue trace).

Importantly, however, significant signal is observed for ion pairs with one of the ions having the mass of the pyrimidine parent ion, i.e., 80 atomic units, emphasized between the red lines in Fig. 1a. To obtain insight into the underlying fragmentation processes, the spectrum of the ions observed *in coincidence* with said parent ion is shown in Fig. 1b.

The most prominent feature in this spectrum is a series of water cluster ions at integer multiple masses of the water molecule ($m/z_{H_2O} = 18$). The slight shift of the peak centers to higher masses compared to the theoretical water cluster ion masses (displayed by dashed vertical bars) may indicate protonation of the water cluster ions, a common observation in the fragmentation of water clusters[22]. Additionally, two peaks are observed at m/z = 1 and m/z = 26-28, which can be attributed to single protons and the pyrimidine fragments $C_2H_2^+$ and $CH_2N^+$. Their appearance is explained by clusters containing more than one pyrimidine molecule, allowing fragmentation into pyrimidine parent ions and molecular fragments.

In order to identify the mechanisms responsible for the experimental observations, we performed calculations on the decay of core vacancies in microsolvated pyrimidine molecules (see Supporting Information for computational details). The electron spectra resulting from the decay of carbon core vacancies of isolated and tetrasolvated pyrimidine are shown in Figs. 2a and b. The spectra are averaged over all carbon atoms of the molecule. The Auger spectrum of isolated pyrimidine (panel 2a) agrees well with earlier experimental and theoretical reports[23]. The total electron spectrum of fourfold hydrated pyrimidine (red trace in panel 2b) is overall very similarly shaped compared to that of isolated pyrimidine. However, a closer look shows a dramatic difference in the nature of the final states. Only a minor part (gray trace in panel 2b) of the spectrum corresponds to dicationic states for which both charges are localized at the pyrimidine molecule. The difference between the total spectrum (red trace) and the localized dicationic contribution corresponds to delocalized final states, for which the second electron was emitted from the environment, having direct implications on the subsequent fragmentation of the system. A very similar behavior is observed for core ionization of nitrogen atoms, and the corresponding spectra are shown in Figs. 2c and d. Exploration of the configurations of the final states reveals core-level intermolecular Coulombic decay[24–26] (core ICD) as the dominant non-local decay (see Supporting Information for a precise description of the contributing channels). In this mechanism, a core vacancy is filled by a valence electron of the pyrimidine, and a valence electron from a neighboring water molecule is emitted. ICD and related processes were recently predicted and observed to significantly contribute to the secondary electron spectra of core and inner-valence vacancies in dense media[27–32].

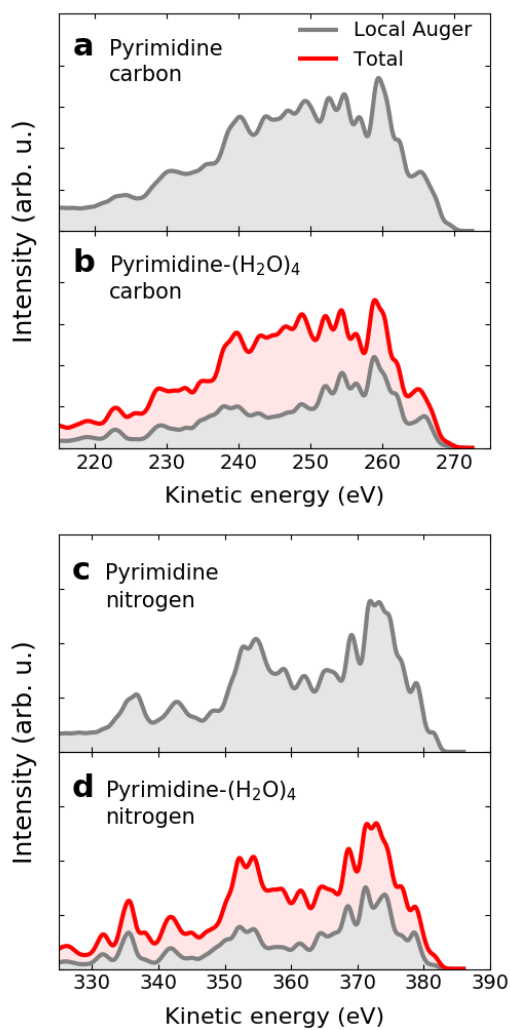

**Figure 2.** Calculated electron spectra resulting from the decay of core vacancies in pyrimidine molecules. **a**, Auger spectrum of isolated pyrimidine after carbon core ionization, averaged over all carbon atoms. **b**, Total spectrum of pyrimidine solvated by four water molecules (red) and the contribution of local Auger decay (gray) after carbon core ionization. **c,** Auger spectrum of isolated pyrimidine after nitrogen core ionization, averaged over both nitrogen atoms. **d,** Total spectrum of pyrimidine solvated by four water molecules (red) and the contribution of local Auger decay (gray) after nitrogen core ionization. In panels **b** and **d**, the red shaded area corresponds to non-local intermolecular channels.

Remarkably, while ICD usually dominates only if local electron emission is energetically forbidden, in the present case it outruns even Auger decay, which is typically the dominant decay route of core vacancies. As one possible consequence of ICD, the pyrimidine molecule and the hydration shell separate by Coulomb repulsion and are observable as pyrimidine parent ion and water cluster ion, matching well with the experimental observation.

For a pyrimidine molecule solvated by only four water molecules, our calculations predict a remarkable ratio of 58 % of carbon core vacancies decaying by intermolecular processes. This value grows from 0 % to 24 %, 41 %, and 50 % for solvation by 0, 1, 2, and 3 water molecules, respectively (see Supporting Information for ionization of nitrogen) and is expected to be even larger for fully solvated molecules.

From the present experimental spectra, the ratio of local to intermolecular decay cannot be deduced, since the exact composition of the target gas jet (isolated pyrimidine, water, pyrimidine molecules hydrated by different numbers of water molecules) is not known accurately. For similar reasons, the exact fraction of molecules, which are protected from fragmentation by the intermolecular decay, is unknown. While some of the final states may still be dissociative (producing pyrimidine fragment ions and neutral fragments), the stable parent ion is known to dominate the mass spectrum after valence ionization [20; 33]. We envision further final state selective studies to quantify the extent of the protection. The protection effect of the solvation against X-ray-induced radiation damage is illustrated in Fig. 3, which schematically shows how different the fates of isolated pyrimidine (panel 3a) and solvated (panel 3b) pyrimidine are upon inner-shell ionization.

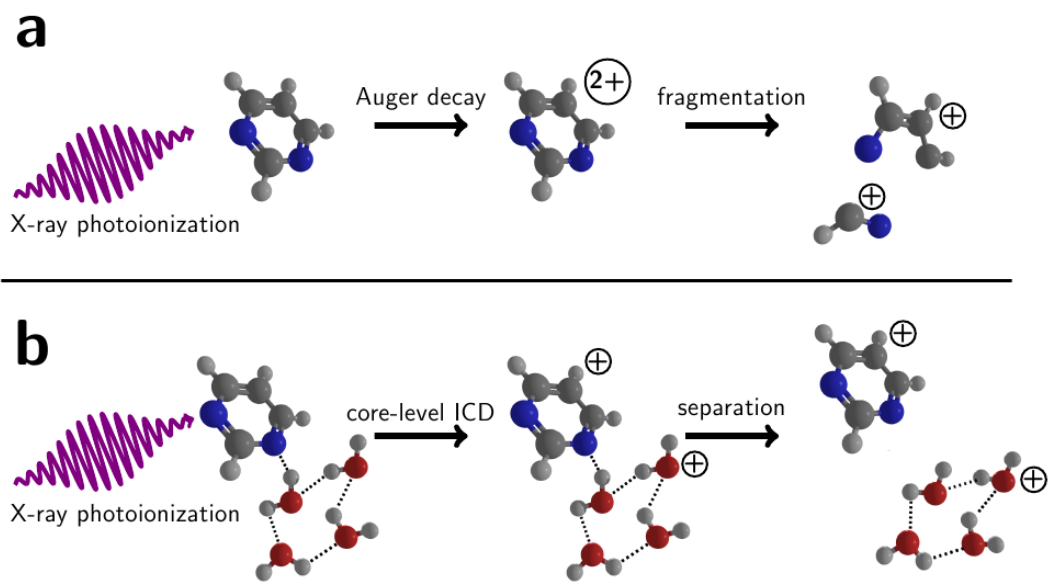

**Figure 3.** Schematic of the observed process. **a,** X-ray photoionization of a biomolecule is followed by Auger decay into doubly charged molecular states. As a consequence, the molecule dissociates into ionic fragments. **b**, If the molecule is solvated, intermolecular decay processes like core-level ICD become operable which distribute charge to the environment and thereby protect the molecule from fragmentation.

The relevance of ICD for radiobiology has been discussed intensely throughout the years after its prediction in 1997[34–37]. Within this discussion, mainly the role of emerging secondary electrons, ions, and radicals contributing to the indirect radiation damage is considered. These particles with damaging potential are the products of ICD of inner-valence vacancies[29; 34] or if ICD is part of a decay cascade and occurs subsequently to local Auger decay[35; 36]. In the present work, we reveal the importance of the core-level variant of ICD as damage protector by demonstrating its decisive role for the fate of a core-ionized biomolecule reducing the direct damage substantially. Our findings imply that models on radiation damage need to be revisited if the damage is mediated by dissociation of multiply charged molecular states, which are populated via excited intermediates. First, the aqueous environment of a bio-relevant molecule starts to intervene early and may drastically change the decay route of excited intermediates before reaching the dissociative state. Second, the products of individual photon-molecule interactions, such as electrons, ions, and radicals, differ qualitatively. Since these products and their properties are considered as the starting conditions for indirect radiation damage, the referring models are also directly affected.

For proper assessment of the radiobiological effect of X-rays, knowledge about the radiation chemistry of ionized solvated molecules is required. X-ray-based radiotherapy mainly relies on the direct damaging effect to biomolecules. Ion radiotherapy turned out to be significantly advantageous compared to X-ray radiotherapy in many points, with the main reason being the lateral concentration of radiation dose[38]. It is another interesting aspect that destructive multiply charged molecular states are typically directly induced by ion impact, while they are populated via excited intermediates after X-ray ionization. Intermolecular decay processes as observed here may thus contribute to the lower efficiency of X-ray radiation therapies.

EXPERIMENTAL METHODS

The experiment was performed at the P04 beamline of the PETRA III synchrotron radiation facility in Hamburg[39]. The storage ring was operated in 40-bunch mode, delivering light pulses to the beamline with about 192 ns temporal spacing. A photoelectron-photoion-coincidence (PEPICO) setup available at the beamline was used. The working principle is similar to the one described in Ref.[40]. The electron spectrometer is a magnetic bottle type time-of-flight spectrometer, the drift tube is equipped with several retardation stages. The permanent magnet, which is used in this type of spectrometers to guide the electrons into the electron drift tube, is ring-shaped and simultaneously is part of the 23 mm long ion drift tube of the ion time-of-flight spectrometer. The ion extraction potential is applied continuously. Therefore, a continuous stable operation without voltage pulsing is possible, however, with the drawback of a limited resolution in both electron and ion spectra. From our ion time-of-flight spectra we estimated a resolution of $(m/z) / \Delta(m/z) \approx 15$. In the electron spectra, photoelectrons and Auger electrons, which are by far the dominant contribution at the used photon energies, are well separated. No further differentiation was attempted. Both ion and electron spectrometer are equipped with a chevron microchannel plate stack to detect electron and ion emission.

Microhydrated pyrimidine molecules were produced by supersonic co-expansion of the vapor of a mixture of liquid water and pyrimidine (94 % water, 6 % pyrimidine) through a conical copper nozzle (80 µm diameter, 30° opening angle) into vacuum. To increase the vapor pressure, the liquid mixture was heated to 80°C. The formation of hydrated molecules under these conditions was confirmed prior to the beamtime in preparatory experiments using a commercial quadrupole mass spectrometer. The ratio of microhydrated molecules is typically in the order of a few percent of the overall target jet. Beside the production of single pyrimidine molecules embedded in a water cluster, the formation of clusters with several pyrimidine molecules is possible but expected to be weak. Note that using this procedure of sample preparation, the appearance of isolated, gaseous water and pyrimidine molecules cannot be avoided.

The expansion chamber was separated from the interaction chamber by a skimmer with 0.7 mm orifice diameter. The typical pressure inside the interaction chamber during operation was about $3 \times 10^{-6}$ mbar.

ACKNOWLEDGMENT

We acknowledge DESY (Hamburg, Germany), a member of the Helmholtz Association HGF, for the provision of experimental facilities. Parts of this research were carried out at PETRA III and we would like to thank the P04 team for assistance in using the P04 beamline and the photoelectron-photoion coincidence spectrometer. This work was supported by the Deutsche Forschungsgemeinschaft (*via* Research Unit FOR 1789 and Project No. 328961117 – SFB 1319 ELCH). L.S.C. is grateful to the European Research Council (Advanced Investigator Grant No. 692657) for financial support.

# Supporting Information associated to manuscript "Suppression of X-Ray-Induced Radiation Damage to Biomolecules in Aqueous Environments by Immediate Intermolecular Decay of Inner-Shell Vacancies"

**Data acquisition and analysis**

The spectrometer settings were adjusted in such a way, that the electrons' time-of-flight distribution was well within the distance of two exciting photon pulses (even for very low kinetic energy electrons). I.e., no electrons were detected later than 192 ns after the respective excitation but features from photoelectrons and Auger electrons were still well separated in the spectra. In contrast, the time-of-flight of the ions considerably exceeded the time interval between two consecutive light pulses. In order to obtain the ion spectra, the ions were correlated to electron coincidences and thereby unambiguously attributed to one exciting pulse. The count rates were chosen such that the contribution from random coincidences was negligible. To assure the negligibility of random coincidences, they were identified as weak features repeating periodically in the time-of-flight ion-ion coincidence presentation. Conversion to m/z ratios of these features confirmed that they do not contribute to the features of interest.

For measurements of photoelectron-photoion-photoion coincidence spectra (PEPIPICO) at the carbon 1s edge, the photon energy of 300 eV (data shown in Fig. 1) was chosen, respectively. At these photon energies, the C1s photoelectrons are slow compared to valence and Auger electrons and are well separated in the time-of-flight spectra.

The calibration of the ion mass-to-charge axis was done using the pyrimidine fragment spectrum present in all measurements and confirmed using the ionization of noble gases.

**Computational details**

*Geometry optimization*. In the present work we considered microhydrated pyrimidine molecules with up to four waters attached and, for comparison, also an isolated pyrimidine molecule. The geometries of all systems were optimized prior exploring their Auger spectra.

The optimized geometries were obtained using the Møller-Plesset perturbation theory in conjunction with the cc-pVTZ[2] basis set (MP2/cc-pVTZ). In agreement with previous studies[1], we found that water molecules in the minimum-energy structures tend to form a subcluster connected to one of the pyrimidine's nitrogen atoms ($N_1$) through a donating hydrogen bond.

*Calculations of the Auger spectra*. The kinetic energies of the emitted Auger electrons were obtained by subtracting the computed double ionization potentials (DIPs) from the binding energies of core electrons. The latter were calculated by means of the ΔSCF/cc-pVDZ method[3].

The number of dicationic states in an isolated pyrimidine molecule is already very large. Yet in microhydrated systems, it becomes tremendous as the vacancies may now be created on different parts of the clusters. To compute all the corresponding DIPs efficiently, we used the second-order algebraic diagrammatic construction method (ADC(2)/cc-pVDZ)[4; 5]. Apart from DIPs, the ADC(2) also yields the transition moments between the ground electronic state and the final dicationic ones.

These are the key elements in the Auger spectrum construction scheme used in our work. By using the two-hole population analysis[6], each transition moment was first decomposed into various contributions representing different combinations of the double vacancies which were then combined in four large classes attributed to the Auger, ICD, ETMD2 and ETMD3 processes. If two final vacancies were in the pyrimidine molecule, then the corresponding contribution was considered a local, Auger contribution. The contributions with two vacancies located in the water sub-cluster were assigned to the ETMD processes and those with a vacancy in pyrimidine and the other one in water were connected to the ICD processes. The character of each dicationic state was thus determined (see different contributions in Figs. 2b and 2d and Extended Data 1).

The absolute electronic decay rates are out of reach for such an extremely high number of decay channels present in the systems studied. However, they are generally not needed, and in order to calculate the spectral intensities, the relative rates are quite enough[6]. The latter were approximated by the local Auger contributions in the transition moments which have double vacancies in an atom subject to core ionization. This method has proved its high efficiency in simulating electronic decay spectra of various molecules where only local decay processes contribute [Refs. 27, 43 and references therein] as well as clusters where local and nonlocal [Refs. 15, 17, 18 and references therein] or even pure nonlocal processes compete[7]. Finally, each computed discrete spectral line was convoluted with a Gaussian of full width at half maximum of 1.5 eV. The C 1s and N 1s Auger spectra shown in Fig. 2 were obtained by averaging the area-normalized spectra of individual atoms.

**Individual channels contributing to the decay of core vacancies in carbon and nitrogen**

The individual channels contributing to the complete non-local spectra are shown in Fig. S1. The structure of the investigated system and the number of solvation molecules are shown as insets. Exemplary, the spectra of one carbon and one nitrogen atom are shown (respective atoms highlighted yellow in the insets). However, due to the delocalization of contributing states, the spectra look similar for all other C and N atoms (not shown). For each individual spectrum, the contribution of local Auger decay (gray shaded spectra) and different intermolecular channels (colored traces) are depicted. In intermolecular Coulombic decay (ICD, red trace), the core vacancy is filled by a valence electron of the pyrimidine molecule and a valence electron from a neighboring water molecule is emitted. In electron-transfer-mediated decay (ETMD), the core vacancy is filled by a valence electron from a neighboring water molecule. In ETMD2 (two molecules involved, blue trace), another valence electron from the same water molecule is emitted, while in ETMD3 (three molecules involved, green trace) yet another adjacent water molecule is ionized. For the investigated scenarios, ICD is the dominant intermolecular channel, in agreement with the experimental observation (pyrimidine parent ions and water cluster ions in coincidence).

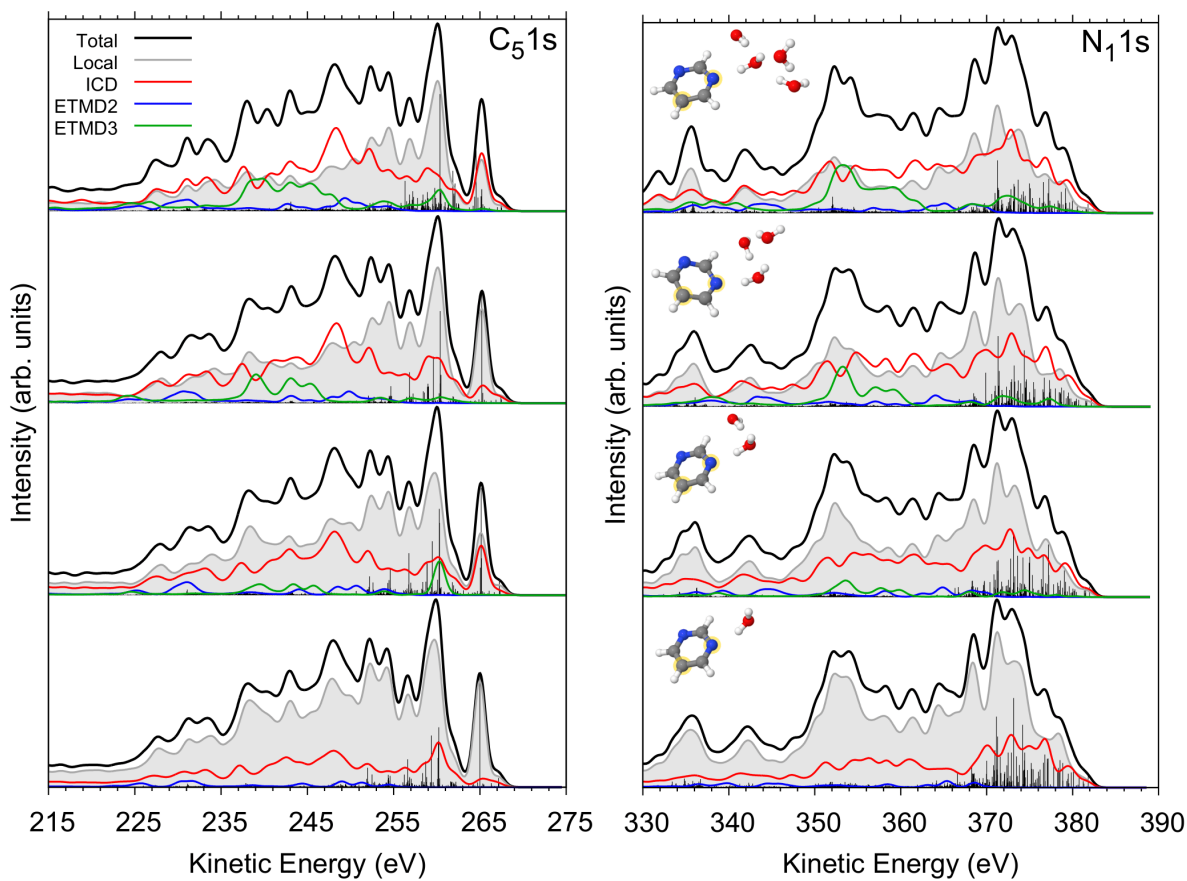

**Figure S1.** Individual channels contributing to the decay of core vacancies in carbon (C, left column) and nitrogen (N, right column) atoms in microsolvated pyrimidine molecules.

The percentage of vacancies which still decays by conventional Auger decay is listed in Tab. S1 for microsolvation of the pyrimidine molecule by 0, 1, 2, 3, and 4 water molecules. For solvation by four water molecules, the integrated intensity of non-local channels outruns Auger decay.

**Table S1.** Calculated contribution (in %) of local Auger decay to the total decay of a core vacancy in a pyrimidine molecule solvated by different numbers of water molecules. The values are averages over all atoms of the respective element in the molecule. The geometric arrangement of the water molecules relative to the pyrimidine is shown in Fig. S1.

| Number of water molecules | C 1s | N 1s |
|---|---|---|
| 4 | 42 | 41 |
| 3 | 50 | 48 |
| 2 | 59 | 58 |
| 1 | 76 | 74 |
| 0 | 100 | 100 |